\documentclass[RNAAS]{aastex63}

\newcommand{\Teff}{$T_\mathrm{eff}$}
\newcommand{\logg}{$\log g$}

\begin{document}

\title{Coadded Spectroscopic Stellar Parameters and Abundances from the LAMOST Low Resolution Survey}

\correspondingauthor{Jacob H. Hamer}\email{jhamer3@jhu.edu}

\author[0000-0002-7993-4214]{Jacob H. Hamer}
\affiliation{Department of Physics and Astronomy, Johns Hopkins University, 3400 N. Charles Street, Baltimore, MD 21218, USA}


\begin{abstract}
    I combine duplicate spectroscopic stellar parameter estimates in the LAMOST Data Release 6 Low Resolution Spectral Survey A, F, G, and K Type stellar parameter catalog. Combining repeat measurements results in a factor of two improvement in the precision of the spectroscopic stellar parameter estimates. Moreover, this trivializes the process of performing coordinate-based cross-matching with other catalogs. Similarly, I combine duplicate stellar abundance estimates for the \citet{Xiang2019} catalog which was produced using LAMOST Data Release 5 Low Resolution Spectral Survey data. These data have numerous applications in stellar, galactic, and exoplanet astronomy. The catalogs I produce are available as machine-readable tables at \url{https://doi.org/10.7281/T1/QISGRU}.
\end{abstract}
\section{} 

The Large Sky Area Multi-Object Fiber Spectroscopic Telescope (LAMOST) Data Release 6 (DR6) contains a catalog of spectroscopic stellar parameters for 5,773,552 A, F, G, and K stars generated from the R$\sim$1800 Low Resolution Survey (LRS) \citep{LAMOSTa, LAMOSTb}. While the Medium Resolution Survey (MRS) data release combines repeat observations into coadded spectra, from which they produce stellar parameter estimates, the LRS does not do so. In this research note I describe the production of a catalogue which combines parameter estimates for objects with repeat observations. 

The LAMOST LRS input catalog describes \texttt{tname} as a unique identifier for each object in the catalog. However, matching the catalog to itself on-sky reveals matches between objects with different \texttt{tname} as well as \texttt{tsource}, with similar spectroscopic stellar parameters. Therefore, \texttt{tname} is not a sufficient identifier to find all parameter estimates for the same star. Additionally, I find that a number of objects from the PILOT survey for which a single \texttt{tname} corresponds to multiple objects with similar declinations but drastically different right ascensions. I therefore undertake the following steps to combine parameter estimates for different observations of the same star. First, I group the catalog by \texttt{tname} and \texttt{tsource}. Then, I check that the positions of the grouped objects do not differ by more than 10 arcseconds in order to exclude the sources from the PILOT survey that have identical \texttt{tname} but different on-sky positions. Next, I check that the radial velocity estimates are not discrepant, only combining measurements when the significance
\begin{equation}
    t = \frac{\mathrm{RV}_{\mathrm{max}} - \mathrm{RV}_{\mathrm{min}}}{\sqrt{\sigma_\mathrm{RV,max}^2+\sigma_\mathrm{RV,min}^2}} 
\end{equation} is less than 3. Such a discrepancy could indicate that the objects are being incorrectly grouped by our method, or that the star is in an unresolved binary. 

For grouped objects which pass the two tests described, I combine their stellar parameter estimates as follows. I calculate the arithmetic weighted mean of each parameter using the SNR in the SDSS g band as the weight: 
\begin{equation}
    \bar{x}=\frac{\sum_i^N \mathrm{SNR}_ix_i}{\sum_i^N \mathrm{SNR}_i},
\end{equation}
where $x_i= \{T_\mathrm{eff}, \mathrm{[Fe/H]}, \log{g}, \mathrm{RV}\}$ denotes a set of spectroscopic stellar parameters from one observation, and SNR is the signal-to-noise ratio in the SDSS g band. I also produce a right ascension and declination for the combined observation with the same equation. It follows from error propagation that the uncertainty on our new parameter estimate is 
\begin{equation}
    \sigma_{\bar{x}}=\frac{\sqrt{\sum_i^N \mathrm{SNR}_i^2 \sigma_{x,i}^2}}{\sum_i^N \mathrm{SNR}_i}.
\end{equation} I estimate the combined SDSS g band SNR of this ``coadded'' observation by adding the g band SNR of each observation in quadrature. I also report the number of observations which go into the new parameter estimate. 

At this point, I have combined observations which can be grouped based on their shared \texttt{tname} and \texttt{tsource}, but have not yet grouped objects which differ in these identifiers but are co-located on the sky. To identify stars which should be combined based on their distance on-sky, I match the grouped catalog with itself on the sky using TOPCAT. I use a radius of 0.5 arcseconds, and I retain all matches. I use the \texttt{GroupID} outputted by TOPCAT as a key to identify which observations to group, and combine them using the same methods described above, including the check that the radial velocities do not have a discrepancy with a significance over 3. 

The catalog which results from these steps to combine multiple observations is 4,334,538 objects. 22\% of these sets of spectroscopic stellar parameters are the result of coadding. 72\% are the result of 2 coadded observations and 28\% are the result of 3 or more. In addition, I produce a catalog which provides sets of \texttt{obsid} which have I determined to correspond to a single object at the listed right ascension and declination. This key allows for one to carry out their own method of coadding repeat observations of an object.

We apply a similar methodology to the catalog of stellar abundances produced by \citet{Xiang2019} using the LAMOST Data Release 5 LRS. I remove any observations which are flagged as poor quality based on the chi-squared of the spectral fit. Additionally, for each observation I use the individual element quality flags  to remove unreliable abundance estimates.  As the [$\alpha$/Fe] estimate is the weighted mean of [Mg/Fe], [Si/Fe], [Ca/Fe], and [Ti/Fe], I remove this estimate if any of the elements are flagged as unreliable. I also omit unreliable estimates of \Teff, \logg, and microturbulent velocity $V_t$. I combine the reliable parameter estimates to produce a weighted mean and associated error as above. The resulting catalog contains 5,792,830 objects, 20\% of which are the result of coadding multiple observations.

\acknowledgements
This material is based upon work supported by the National Science Foundation under grant number 2009415.

\bibliography{bib}

\end{document}